\documentclass{article}
\usepackage[square,numbers]{natbib}
\usepackage{authblk}

\usepackage{graphicx} % Required for inserting images
\usepackage{comment}
\usepackage{float}
\usepackage{makecell}
\usepackage{amsmath}
\usepackage{tablefootnote}
\usepackage{subcaption}
\usepackage{cleveref}
\usepackage{url}

\providecommand{\keywords}[1]
{
  \small	
  \textbf{\textit{Keywords---}} #1
}

\begin{document}

\title{Uncovering Regional Defaults from Photorealistic Forests in Text-to-Image Generation with DALL·E 2} 

\author[1]{Zilong Liu}
\author[1]{Krzysztof Janowicz}
\author[2]{Kitty Currier}
\author[1]{Meilin Shi}

\affil[1]{Department of Geography and Regional Research, University of Vienna, Austria}
\affil[2]{Department of Geography, University of California, Santa Barbara, USA}
\date{}

\maketitle

\begin{abstract}
\textit{Regional defaults} describe the emerging phenomenon that text-to-image (T2I) foundation models used in generative AI are prone to over-proportionally depicting certain geographic regions to the exclusion of others. In this work, we introduce a scalable evaluation for uncovering such regional defaults. The evaluation consists of region hierarchy--based image generation and cross-level similarity comparisons. We carry out an experiment by prompting DALL·E 2, a state-of-the-art T2I generation model capable of generating photorealistic images, to depict a forest. We select \emph{forest} as an object class that displays regional variation and can be characterized using spatial statistics. For a region in the hierarchy, our experiment reveals the regional defaults implicit in DALL·E 2, along with their scale-dependent nature and spatial relationships. In addition, we discover that the implicit defaults do not necessarily correspond to the most widely forested regions in reality. Our findings underscore a need for further investigation into the geography of T2I generation and other forms of generative AI.
\end{abstract}

\section{Introduction}
\label{sec:intro}
Recent breakthroughs in generative AI have attracted researchers to explore how foundation models represent geographic space. Most recent work focuses on large language models (LLMs), studying how to efficiently extract rich, geographic information~\cite{roberts2023gpt4geo,manvi2023geollm} from them as knowledge bases~\cite{petroni2019language}. However, LLMs face a geographical bias which manifests as a lack of knowledge about some regions (or places) compared with others \cite{janowicz2023philosophical}. For instance, certain parts of the world, namely the US and the UK, may be represented better than South and Southeast Asia~\cite{dunn2024pre}. Disparities also become evident through geoparsing~\cite{mai2023opportunities}, and knowledge probing about geographic features~\cite{liu2024measuring} and common sense shared by local populations~\cite{yin2022geomlama}.

In this work, our focus is on text-to-image (T2I) generation models. A T2I generation model is a type of multimodal LLM that generates images based on a user's prompt in natural language. In this paper, we aim to observe how one such model represents space by prompting it to depict an object class that varies geographically. A recent community-centered study revealed that T2I generation may amplify cultural defaults, favoring cultural subjects from particular hegemonic cultures in their depictions~\cite{qadri2023ai}. For instance, besides over-representing Western-centric subjects, T2I models tended to picture more Indian subjects than Pakistani or Bangladeshi subjects. This made the West a \textbf{regional} cultural default for the world, and India a default for South Asia. Moreover, if observing the default for the world at a country level, will the West become the US? Due to the potential scale dependency of regional defaults, prompts rich in both spatial coverage and granularity are required for the purpose of evaluating T2I generation.

If a T2I generation model is trained on geographically unbalanced data, regions with larger amounts of data are likely to be over-represented in its output. This unintended consequence may arise for many object classes, causing instances from over-represented regions to be inadvertently adopted as canonical representations of their class, i.e. regional defaults. In this work, we choose \emph{forest} as the class of interest, as forests vary geographically in appearance (and other characteristics, of course), and they have a measurable spatial extent when clearly defined%\footnote{While many different definitions of \emph{forest} are in use~\cite{lund2002forest}, for our purposes, selecting a source that reported forest statistics in a globally consistent manner was most important.}
. According to the Global Forest Resources Assessment 2020 (FRA 2020)\footnote{\url{https://www.fao.org/documents/card/en/c/ca9825en}}, forests cover an area of 4.06 billion hectares. This number is equivalent to 31\% of Earth's land surface, making forests a significant global ecosystem. Due to pressing concerns such as climate change and deforestation, many efforts have been made to monitor changes in forest cover through remote sensing. We can use available spatial statistics, such as the distribution and extent, to examine the association between real-world forests and their AI-generated image counterparts.

More concretely, our work tries to understand the generated geography of DALL·E 2\footnote{\url{https://openai.com/dall-e-2}}, a T2I model by OpenAI that can generate photorealistic images. Its successor, DALL·E 3, tends to generate hyper-realistic images and is therefore not considered in this study. Our work does not involve the use of photographs. We leave the assessment of how closely generated images match photographs taken in a particular geographic region for future work. Rather, we argue that valuable insights about the geography reflected in the output of a model can be derived from the underlying spatial relationships within a set of generated data, alone. By prompting with a region-hierarchy dataset, we first obtain forest images that span the globe at different scales. We discover regional defaults from photorealistic forests based on which region in the hierarchy an image resembles the most. To achieve this goal, we apply two image similarity measures. Additionally, we examine both the frequency distribution of regional-default degree measured as the similarity between a regional default and its higher-level region. Lastly, we compare whether regional defaults match their real-world counterparts, given the spatial distribution and extent of forests in reality.

The outline of the paper is structured as follows. We describe the methods used in Section~\ref{sec:methods} and information about generated images in Section~\ref{sec:img_info}. The analysis results are included in Section~\ref{sec:results}. In Section~\ref{sec:conclusions}, we conclude our study and propose future work. 

\section{Methods}
\label{sec:methods}

\subsection{Region-based Forest Image Generation}
\label{sec:generation}
For forest image generation, we refer to the ISO 3166 Country and Dependent Territories Lists with UN Regional Codes\footnote{\url{https://github.com/lukes/ISO-3166-Countries-with-Regional-Codes}}. This source includes names of countries and dependent territories in the ISO 3166 standard, aggregated into regions, sub-regions, and intermediate regions defined by the UN Statistics Division. Using this data source, we build a region hierarchy structured as \texttt{World -> UN region -> UN sub-region -> UN intermediate region -> ISO country or dependent territory}. Not all UN sub-regions have UN intermediate regions, and not all ISO countries or dependent territories have associated UN intermediate regions. We write prompts such as \texttt{forest in Asia} to generate one forest image per region. When generating the image for \texttt{World}, however, we prompt only with \texttt{forest}. When prompting DALL·E 2, the size of generated images is set to be \texttt{1024x1024 pixels} and each prompt is submitted in a separate session.

\subsection{Cross-level Image Similarity Comparison}
\label{sec:imgsim}
In order to discover regional defaults from photorealistic forests, we compute a cross-level image similarity. The cross-level image similarity is defined as the similarity between an image generated for a higher-level region (e.g., \texttt{World}) and an image generated for a lower-level region (e.g., \texttt{Austria}) that belongs to the higher-level region. Therefore, the regional default for a higher-level region is determined by its region with the highest cross-level similarity at a specific lower level. Before computing the similarity, we convert generated images from RGB to grayscale color space. Then, we apply two image similarity measures. The image processing is implemented with OpenCV\footnote{\url{https://opencv.org}} and scikit-image\footnote{\url{https://scikit-image.org}}.

The first measure is the mean squared error (MSE), which measures the average squared difference between pixel values of two images. While MSE is a simple implementation of similarity computation, it works well when there is no significant transformation between two generated images. We consider this possible outcome in DALL·E 2 generation. A smaller MSE represents a higher image similarity. In Equation~\ref{eq:mse}, $x_i$ and $y_i$ are the corresponding values of the $i^{th}$ pixel in a pair of images; $n$ stands for the total number of pixels.
\begin{equation}
\label{eq:mse}
    MSE=\frac{1}{n}\sum_{i=1}^{n}(x_i-y_i)^2
\end{equation}
The second measure is the structural similarity index measure (SSIM)~\cite{wang2004image}. SSIM provides results based on human visual perception by considering luminance, contrast, and structure. A higher SSIM means a higher similarity. In Equation~\ref{sec:ssim}, $x$ and $y$ are two corresponding windows for a pair of images, and the windows are used for calculating local statistics; $l(x, y)$, $c(x, y)$, and $s(x, y)$ are the comparison functions for luminance, contrast, and structure, respectively; $\alpha$, $\beta$, and $\gamma$ are weights for comparison functions; $\mu_x$ (or $\mu_y$) is the mean of pixel values of $x$ (or $y$); $\sigma_x$ (or $\sigma_y$) is the standard deviation of pixel values of $x$ (or $y$); $\sigma_{xy}$ is the covariance between pixel values of $x$ and $y$; $c_1$ and $c_2$ are two constants that prevent division by zero, and $c_3$ is a constant set to be $\frac{1}{2}c_2$. The SSIM between two images is the average $SSIM(x,y)$. We use the default parameter settings of scikit-image for the window size, weights, and constants in the computation.
\begin{equation}
\label{sec:ssim}
 \begin{aligned}
    SSIM(x, y) &= l(x, y)^\alpha \cdot c(x, y)^\beta \cdot s(x, y)^\gamma \\
    l(x, y) &= \frac{2\mu_x \mu_y + c_1}{\mu_x^2 + \mu_y^2 + c_1} \\
    c(x, y) &= \frac{2\sigma_x \sigma_y + c_2}{\sigma_x^2 + \sigma_y^2 + c_2} \\
    s(x, y) &= \frac{\sigma_{xy} + c_3}{\sigma_x \sigma_y + c_3}
    \end{aligned}
\end{equation}

\section{Generated Image Information}
\label{sec:img_info}
\begin{figure}[h]
\caption{A subset of forest images generated following the region hierarchy}
\centering
\includegraphics[width=0.75\linewidth]{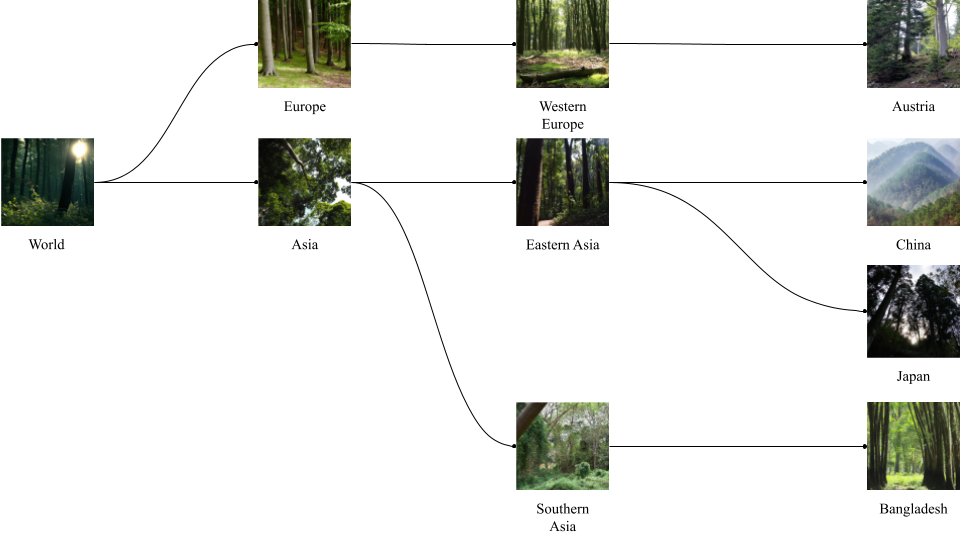}
\label{fig:forests}
\end{figure}

\begin{table}[h]
    \centering
\caption{Number of forest images generated at each level}
\label{tab:image_stat}
\scalebox{0.75}{
    \begin{tabular}{|c|c|c|c|c|c|} \hline 
         \textbf{Level}&  World&  UN region&  UN sub-region & \makecell{UN\\intermediate region}& \makecell{ISO country or\\dependent territory}\\ \hline
 \textbf{Total}& 1& 5& 17& 8& 244\\\hline
    \end{tabular}
    }
\end{table}
Table~\ref{tab:image_stat} shows the total number of forest images generated at each level. Forest images could not be generated for five ISO countries or dependent territories including Eritrea, South Sudan, Sudan, Virgin Islands (British), and Virgin Islands (U.S.), because they were rejected by OpenAI's Safety Systems\footnote{\url{https://openai.com/safety/safety-systems}}. Therefore, these regions are excluded from the subsequent analysis. Figure~\ref{fig:forests} shows a subset of images generated following the region hierarchy, where the level decreases from lefofoft to right.

\section{Results}
\label{sec:results}

\subsection{Regional Defaults for \texttt{World}}
\label{sec:defaults}
Here, we present a case study about regional defaults for the root region, i.e., \texttt{World}. Table~\ref{tab:sim_world} shows the most similar regions, i.e., regional defaults, and the least similar regions to \texttt{World}. Regarding regional defaults, only at the level of UN sub-region did both MSE and SSIM result in the same answer---Latin America and the Caribbean. At any given level, the two measures did not provide the same answer about the least similar region.

We notice a phenomenon that lower-level regional defaults were not necessarily geographically consistent with higher-level regional defaults. For instance, at the lowest level, Nauru was characterized as most similar to \texttt{World}, whereas at the UN-region level, the Americas was most similar to \texttt{World}, based on MSE. As Nauru, a country in Oceania, is distinct from the Americas, this result was geographically inconsistent. The same observations were applicable to regional defaults detected with SSIM, and the least similar regions as well. Regional defaults were most consistent across levels based on MSE, as shown by \texttt{Americas->Latin America and the Caribbean->South America} following the region hierarchy. On the contrary, no single pair of regional defaults followed the hierarchy based on SSIM. Whether based on MSE or SSIM, the least similar regions did not strongly follow the hierarchy, either. For instance, MSE only resulted in \texttt{Africa->Northern Africa}, and SSIM only resulted in \texttt{Oceania->Melanesia}.
\begin{table}[h]
\centering
\caption{The most and least similar regions to \texttt{World} based on MSE and SSIM}
\label{tab:sim_world}
\scalebox{0.75}{
    \begin{tabular}{|c|c|c|c|c|} \hline  
           \textbf{Level}& \textbf{\makecell{Most Similar\\(MSE)}} &\textbf{\makecell{Most Similar\\(SSIM)}}&\textbf{\makecell{Least Similar\\(MSE)}}&\textbf{\makecell{Least Similar\\(SSIM)}}\\ \hline  
           UN region& \makecell{Americas\\(4323)}& \makecell{Europe\\(0.18)}& \makecell{Africa\\(14897)}& \makecell{Oceania\\(0.06)}\\ \hline  
           UN sub-region& \makecell{Latin America\\and the Caribbean\\(3361)} &\makecell{Latin America\\and the Caribbean\\(0.30)}&\makecell{Northern Africa\\(14128)}&\makecell{Melanesia\\(0.07)}\\ \hline  
           \makecell{UN\\intermediate region}& \makecell{South America\\(4983)}&\makecell{Southern Africa\\(0.13)}&\makecell{Eastern Africa\\(13346)}&\makecell{Channel Islands\\(0.07)}\\ \hline  
           \makecell{ISO country or\\dependent territory}& \makecell{Nauru\\(2922)}&\makecell{Uganda\\(0.26)}&\makecell{Antarctica\\(18066)}&\makecell{Sint Maarten\\(Dutch part)\\(0.04)}\\\hline 
    \end{tabular}
    }
\end{table}
\subsection{The Frequency Distribution of Regional-Default Degree}
\begin{figure}[h]
    \centering
    \caption{The frequency distribution of regional-default degree}
     \begin{subfigure}[b]{0.45\linewidth}
         \centering
         \includegraphics[width=0.75\linewidth]{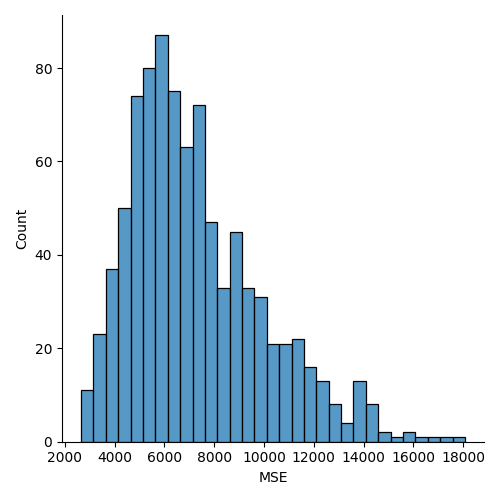}
         \caption{MSE (bin width = 500)}
        \label{fig:prob_mse}
     \end{subfigure}
     \hfill
     \begin{subfigure}[b]{0.45\linewidth}
         \centering
         \includegraphics[width=0.75\linewidth]{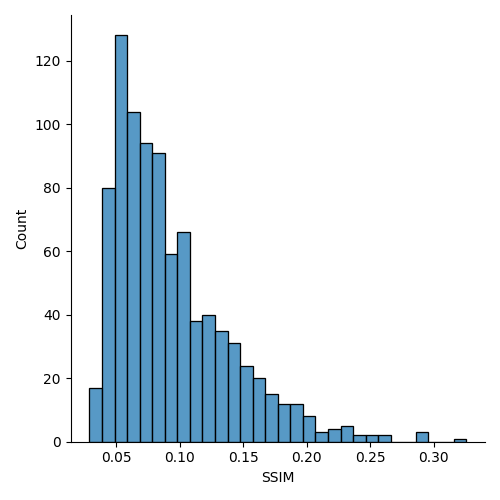}
         \caption{SSIM (bin width = 0.01)}
         \label{fig:prob_ssim}
     \end{subfigure}
     \label{fig:prob}

\end{figure}

All MSE and SSIM values are reported in Table~\ref{tab:sim_world}. As MSE outputs absolute errors, here we focus on SSIM values that have a range of $[-1,1]$. According to our experiment, regions with the strongest regional-default degree---Latin America and the Caribbean---only had an SSIM value of 0.30. This value had no substantial difference compared with the least similar regions at the same level---Melanesia, with a value of 0.07. If we do not want to simply choose the most similar regions, can we obtain a similarity threshold from the statistical pattern of region-default degree? Figure~\ref{fig:prob} shows the frequency distribution of regional-default degree for all observations. Both MSE-based (Figure~\ref{fig:prob_mse}) and SSIM-based (Figure~\ref{fig:prob_ssim}) distributions were right skewed. As a higher SSIM value corresponds to a higher similarity, a threshold could be determined as the cut-off value that separates the right tail from the distribution. The value was approximately 0.21 according to Figure~\ref{fig:prob_ssim}. In Table~\ref{tab:sim_world}, Latin America and the Caribbean (0.30) and Uganda (0.26) had an SSIM value over 0.21. This means that if 0.21 was applied as the threshold, regional defaults for \texttt{World} would exist only at two levels. On the contrary, no cut-off could be chosen from the left tail of the MSE-based distribution. This made SSIM a better image similarity measure than MSE if an appropriate threshold was desired.

\subsection{Regional Defaults as the Most Widely Forested Regions?}
\label{sec:forest_regions}
When we examine Table~\ref{tab:sim_world} again, we notice that Nauru, as a regional default based on MSE, contradicted our common sense because the country was not forested according to FRA 2020. However, Antarctica, as a non-forested continent, was the least similar region to \texttt{World} based on MSE. This observation inspires us to compare whether these regional defaults corresponded to real-world regions with the largest forest extent. To validate our hypothesis, we use forest-extent statistics from the data portal\footnote{\url{https://fra-data.fao.org/assessments/fra/2020}} of FRA 2020 to obtain figures for the most widely forested regions at each level. Table~\ref{tab:largest_region} shows the forest extent of these regions, among which the Americas and South America were also regional defaults detected in Table~\ref{tab:sim_world} based on MSE. This indicates that regional defaults for \texttt{World} at the level of UN region and the level of UN intermediate region matched their real-world spatial statistics. Additionally, we observe that the most widely forest regions at a lower level were also geographically inconsistent with regions at a higher level. This aligned with our conclusion about the spatial relationships among regional defaults in the case study.
\begin{table}[h]
    \centering
\caption{The most widely forested regions at each level based on FRA 2020 statistics}
\label{tab:largest_region}
\scalebox{0.75}{
    \begin{tabular}{|c|c|c|} \hline  
           \textbf{Level}& \textbf{Most Forested Region}&\textbf{Forest Extent ($10^6 km^2$)}\\ \hline  
           UN region& Americas& 13.6\\ \hline  
           UN sub-region& Eastern Europe &8.3\\ \hline  
           \makecell{UN\\intermediate region}& South America&6.1\\ \hline  
           \makecell{ISO country or\\dependent territory}& Russian Federation&8.1\\\hline 
    \end{tabular}
    }
\end{table}

\section{Conclusions and Future Work}
\label{sec:conclusions}
In this work, we identify an emerging geographical issue, namely regional defaults in T2I generation. We characterize regional defaults as regions that are amplified and appear as defaults in generated image sets, and we demonstrate that regional defaults are scale-dependent. Using \emph{forest} as an example object class, we carry out an evaluation of DALL·E 2. We follow a region hierarchy to generate photorealistic forests for a total of 275 regions at five levels. Using two image similarity measures, MSE and SSIM, we compute a cross-level similarity to detect regional defaults. There are three major findings. \textbf{First}, our case study about regional defaults for \texttt{World} shows that lower-level defaults are not necessarily geographically consistent with higher-level defaults. \textbf{Second}, the frequency distribution of regional-default degree for all observations indicates the advantage of SSIM over MSE when a similarity threshold is desired. \textbf{Lastly}, the analysis using forest-extent data from FRA 2020 reveals that the observed regional defaults do not necessarily correspond to regions with the largest forest extent. Our evaluation can be extended with any region hierarchy, but how susceptible T2I generation is to place-name ambiguity~\cite{leidner2008toponym} in a prompt requires more attention. Our findings once again demonstrate the value in building geographic information observatories~\cite{janowicz2023philosophical} to investigate how geography---both real and manipulated, as in the case of deepfake geography~\cite{zhao2021deep}---are represented by generative AI. Such observatories should investigate more images per region, object classes, models, and advanced similarity measures.

\bibliography{reference}

\begin{thebibliography}{12}
\providecommand{\natexlab}[1]{#1}
\providecommand{\url}[1]{\texttt{#1}}
\expandafter\ifx\csname urlstyle\endcsname\relax
  \providecommand{\doi}[1]{doi: #1}\else
  \providecommand{\doi}{doi: \begingroup \urlstyle{rm}\Url}\fi

\bibitem[Dunn et~al.(2024)Dunn, Adams, and Madabushi]{dunn2024pre}
J.~Dunn, B.~Adams, and H.~T. Madabushi.
\newblock Pre-trained language models represent some geographic populations better than others.
\newblock \emph{arXiv preprint arXiv:2403.11025}, 2024.

\bibitem[Janowicz(2023)]{janowicz2023philosophical}
K.~Janowicz.
\newblock Philosophical foundations of geoai: Exploring sustainability, diversity, and bias in geoai and spatial data science.
\newblock In \emph{Handbook of Geospatial Artificial Intelligence}, pages 26--42. CRC Press, 2023.

\bibitem[Leidner(2008)]{leidner2008toponym}
J.~L. Leidner.
\newblock \emph{Toponym resolution in text: Annotation, evaluation and applications of spatial grounding of place names}.
\newblock Universal-Publishers, 2008.

\bibitem[Liu et~al.(2024)Liu, Janowicz, Currier, and Shi]{liu2024measuring}
Z.~Liu, K.~Janowicz, K.~Currier, and M.~Shi.
\newblock Measuring geographic diversity of foundation models with a natural language--based geo-guessing experiment on gpt-4.
\newblock \emph{AGILE: GIScience Series}, 5:\penalty0 38, 2024.

\bibitem[Mai et~al.(2023)Mai, Huang, Sun, Song, Mishra, Liu, Gao, Liu, Cong, Hu, et~al.]{mai2023opportunities}
G.~Mai, W.~Huang, J.~Sun, S.~Song, D.~Mishra, N.~Liu, S.~Gao, T.~Liu, G.~Cong, Y.~Hu, et~al.
\newblock On the opportunities and challenges of foundation models for geospatial artificial intelligence.
\newblock \emph{arXiv preprint arXiv:2304.06798}, 2023.

\bibitem[Manvi et~al.(2023)Manvi, Khanna, Mai, Burke, Lobell, and Ermon]{manvi2023geollm}
R.~Manvi, S.~Khanna, G.~Mai, M.~Burke, D.~Lobell, and S.~Ermon.
\newblock Geollm: Extracting geospatial knowledge from large language models.
\newblock \emph{arXiv preprint arXiv:2310.06213}, 2023.

\bibitem[Petroni et~al.(2019)Petroni, Rockt{\"a}schel, Lewis, Bakhtin, Wu, Miller, and Riedel]{petroni2019language}
F.~Petroni, T.~Rockt{\"a}schel, P.~Lewis, A.~Bakhtin, Y.~Wu, A.~H. Miller, and S.~Riedel.
\newblock Language models as knowledge bases?
\newblock \emph{arXiv preprint arXiv:1909.01066}, 2019.

\bibitem[Qadri et~al.(2023)Qadri, Shelby, Bennett, and Denton]{qadri2023ai}
R.~Qadri, R.~Shelby, C.~L. Bennett, and E.~Denton.
\newblock Ai’s regimes of representation: A community-centered study of text-to-image models in south asia.
\newblock In \emph{Proceedings of the 2023 ACM Conference on Fairness, Accountability, and Transparency}, pages 506--517, 2023.

\bibitem[Roberts et~al.(2023)Roberts, L{\"u}ddecke, Das, Han, and Albanie]{roberts2023gpt4geo}
J.~Roberts, T.~L{\"u}ddecke, S.~Das, K.~Han, and S.~Albanie.
\newblock Gpt4geo: How a language model sees the world's geography.
\newblock \emph{arXiv preprint arXiv:2306.00020}, 2023.

\bibitem[Wang et~al.(2004)Wang, Bovik, Sheikh, and Simoncelli]{wang2004image}
Z.~Wang, A.~C. Bovik, H.~R. Sheikh, and E.~P. Simoncelli.
\newblock Image quality assessment: from error visibility to structural similarity.
\newblock \emph{IEEE transactions on image processing}, 13\penalty0 (4):\penalty0 600--612, 2004.

\bibitem[Yin et~al.(2022)Yin, Bansal, Monajatipoor, Li, and Chang]{yin2022geomlama}
D.~Yin, H.~Bansal, M.~Monajatipoor, L.~H. Li, and K.-W. Chang.
\newblock Geomlama: Geo-diverse commonsense probing on multilingual pre-trained language models.
\newblock \emph{arXiv preprint arXiv:2205.12247}, 2022.

\bibitem[Zhao et~al.(2021)Zhao, Zhang, Xu, Sun, and Deng]{zhao2021deep}
B.~Zhao, S.~Zhang, C.~Xu, Y.~Sun, and C.~Deng.
\newblock Deep fake geography? when geospatial data encounter artificial intelligence.
\newblock \emph{Cartography and Geographic Information Science}, 48\penalty0 (4):\penalty0 338--352, 2021.

\end{thebibliography}

\end{document}